\documentclass[twocolumn,showpacs,preprintnumbers,amsmath,amssymb]{revtex4}
\usepackage{graphicx}
\usepackage{dcolumn}
\usepackage{bm}
\begin{document}
\title{Collective depinning and sliding of a quantum Wigner solid in a 2D electron system}
\author{M.~Yu.~Melnikov and A.~A. Shashkin}
\affiliation{Institute of Solid State Physics, Chernogolovka, Moscow District 142432, Russia}
\author{S.-H. Huang and C.~W. Liu}
\affiliation{Department of Electrical Engineering and Graduate Institute of Electronics Engineering, National Taiwan University, Taipei 106, Taiwan}
\author{S.~V. Kravchenko}
\affiliation{Physics Department, Northeastern University, Boston, MA 02115, USA}
\begin{abstract}
We report the observation of two-threshold voltage-current characteristics accompanied by a peak of broadband current noise between the two threshold voltages in the insulating state at low densities in the 2D electron system in ultra-high mobility SiGe/Si/SiGe heterostructures.  The observed results can be described by a phenomenological theory of the collective depinning of elastic structures, which naturally generates a peak of a broadband current noise between the dynamic and static thresholds and changes to sliding of the solid over a pinning barrier above the static threshold.  This gives compelling evidence for the electron solid formation in this electron system and shows the generality of the effect for different classes of electron systems.
\end{abstract}
\maketitle
\setlength{\parskip}{0pt}

Strongly correlated electron systems are at the forefront of condensed matter physics; theoretical methods in this field are still being developed, and experimental results are in great demand.  The strongly interacting regime in two-dimensional (2D) electron systems is realized at low electron densities, where the Coulomb energy strongly exceeds the Fermi energy.  The strength of interactions is characterized by a dimensionless parameter $r_{\rm s}=g_{\rm v}/(\pi n_{\rm s})^{1/2}a_{\rm B}$ (here $g_{\rm v}$ is the valley degeneracy, $n_{\rm s}$ is the electron density and $a_{\rm B}$ is the Bohr radius in semiconductor).  The ground state of a 2D electron system in the strongly interacting limit has been predicted to be a quantum Wigner crystal at $r_{\rm s}\gtrsim35$, which may be preceded by possible intermediate states \cite{chaplik1972possible,tanatar1989ground,attaccalite2002correlation,spivak2004phases,spivak2006transport,zverev2012microscopic,dolgopolov2017quantum}.  Many experimental results on low-temperature transport and thermodynamic properties of 2D electrons in semiconductors indicate a sharp increase of the effective mass and spin susceptibility, proportional to the product of the effective mass and the Land\'e $g$-factor, near the metal-insulator transition \cite{shashkin2001indication,shashkin2002sharp,pudalov2002low,zhu2003spin,tan2005measurements,shashkin2006pauli,gokmen2010contrast,mokashi2012critical,kasahara2012correlation,falson2015electron,kuntsevich2015strongly, melnikov2017indication,melnikov2019quantum,falson2022competing,melnikov2023spin}, with critical behavior reported in Refs.~\cite{shashkin2001indication,fletcher2001critical,shashkin2002sharp,shashkin2006pauli,mokashi2012critical,melnikov2019quantum,melnikov2023spin}.  This points to a phase transition to a new state at low electron densities that can be either a quantum Wigner crystal or a precursor.  Single-threshold current-voltage ($I$-$V$) characteristics were observed in the insulating regime in a number of 2D electron systems (see, \textit{e.g.}, Refs.~\cite{chitra2005zero,knighton2018evidence,falson2022competing,hossain2022anisotropic,madathil2023moving}) and were tentatively interpreted in terms of a Wigner crystal formation.  However, these results could also be explained by mundane mechanisms like percolation \cite{jiang1991magnetotransport,dolgopolov1992metal,shashkin1994insulating} or hopping in a strong electric field \cite{marianer1992effective}.

\begin{figure}[b]
\scalebox{1}{\includegraphics[width=\columnwidth]{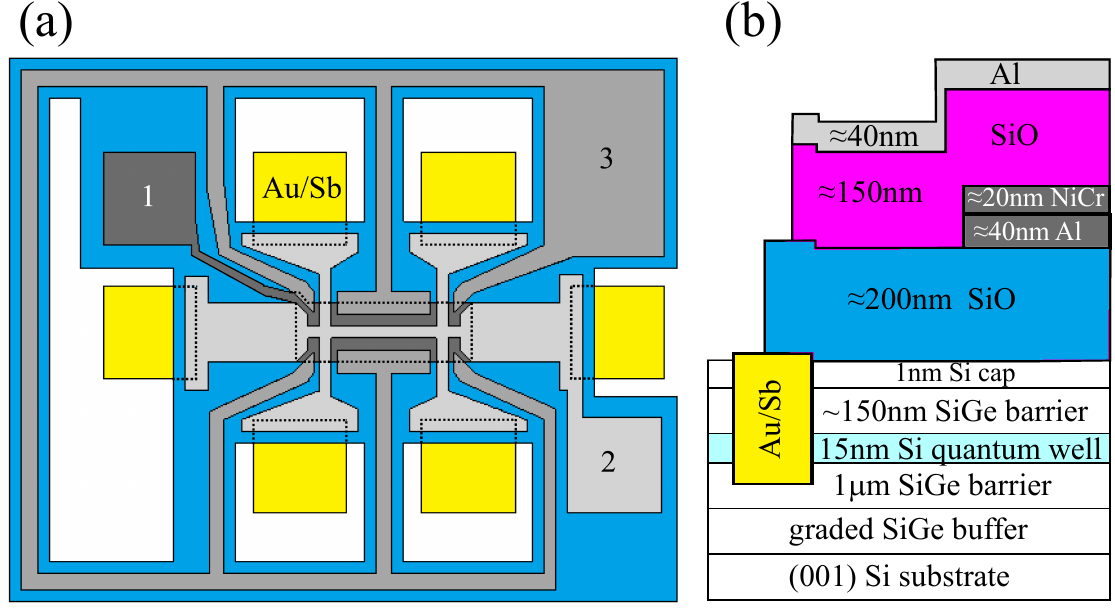}}
\caption{(a) Schematic top view on the sample with three independent gates. The Hall-bar-shaped gate 1 is hidden and is shown by the dashed line in the central part of the sample. Contact gate 2 and depleting gate 3 are above gate 1. (b) Schematics of the layer growth sequence and the cross section of the triple-gate sample.}
\label{fig1}
\end{figure}

\begin{figure*}[t]
\scalebox{.76}{\includegraphics[width=\columnwidth]{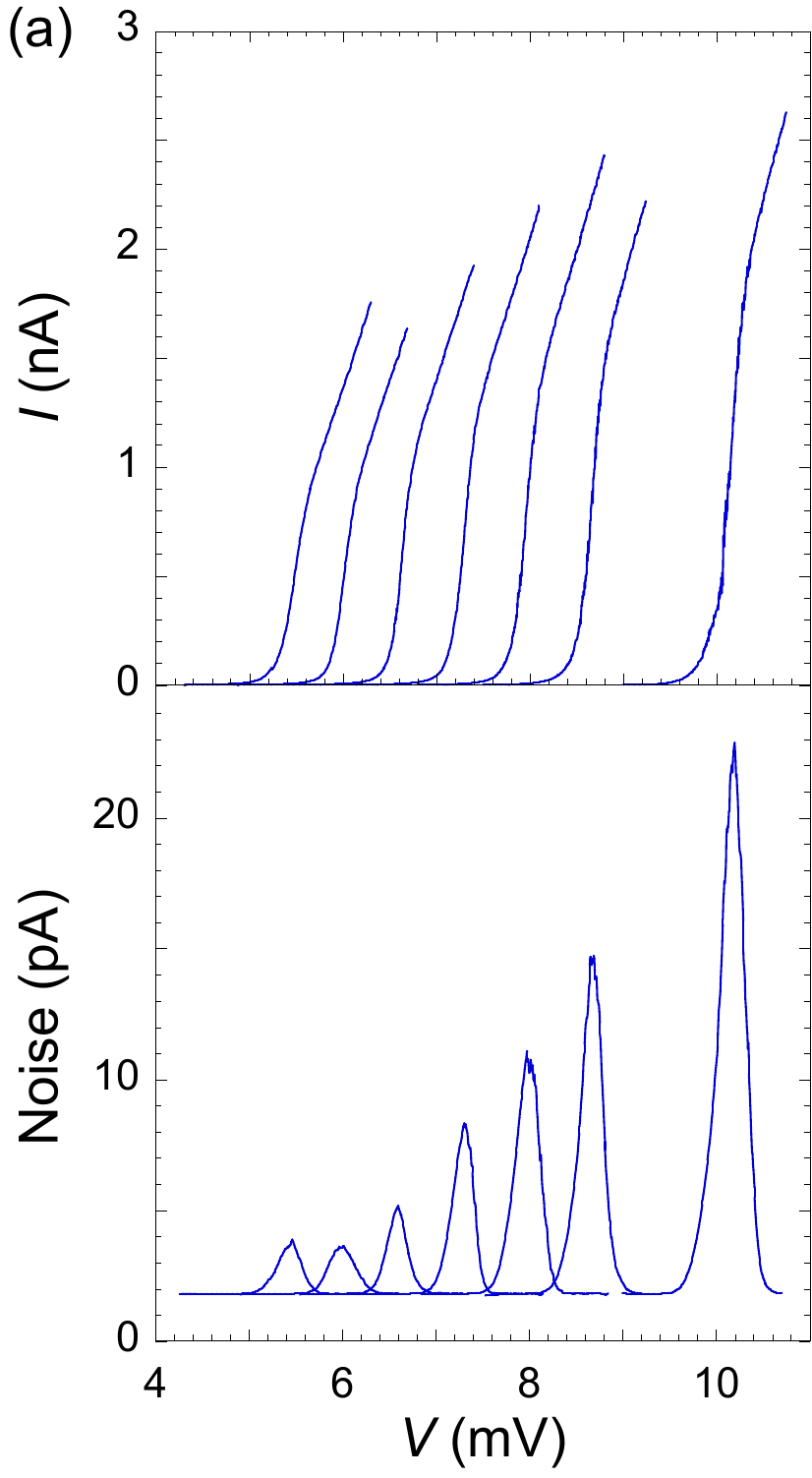}}\hspace{15mm}
\scalebox{.765}{\includegraphics[width=\columnwidth]{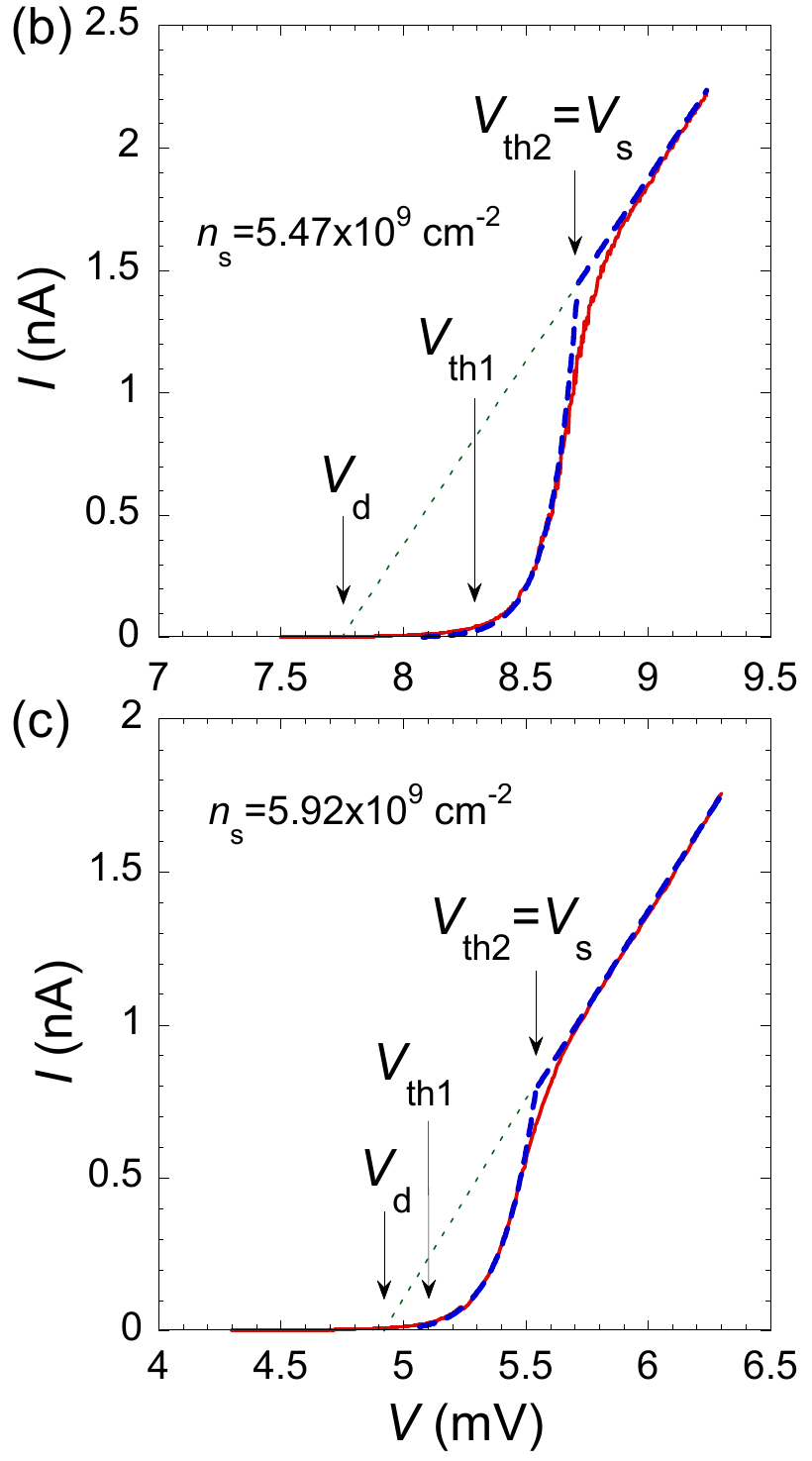}}
\caption{(a) $V$-$I$ characteristics (upper panel) and the broadband noise (lower panel) in sample 1 at a temperature of $\approx30$~mK and different electron densities (left to right): 5.92, 5.83, 5.74, 5.65, 5.56, 5.47, and $5.29\times10^9$~cm$^{-2}$.  (b) and (c) $V$-$I$ characteristics for two electron densities on an expanded scale; also shown are the threshold voltages, $V_{\rm{th1}}$ and $V_{\rm{th2}}$, the dynamic threshold $V_{\rm d}$ obtained by the extrapolation of the linear part of the $V$-$I$ curves to zero current, and the static threshold $V_{\rm s}=V_{\rm{th2}}$.  The dashed lines are fits to the data using Eqs.~(\ref{1}), (\ref{2}).}
\label{fig2}
\end{figure*}

Recently, two-threshold $V$-$I$ characteristics that reveal the signature of a Wigner solid and exclude mundane interpretations in terms of percolation or overheating have been observed in high-mobility silicon metal-oxide-semiconductor field-effect transistors (MOSFETs) \cite{brussarski2018transport} where the disorder has a short-range character caused by the residual point scatterers at the interface, and the critical density for the interaction-induced metal-insulator transition ($n_{\rm c}\approx8\times10^{10}$~cm$^{-2}$) is relatively high.  This is in contrast to the single-threshold $V$-$I$ characteristics observed in heterostructures with the carrier mobility higher by orders of magnitude, including electron and hole GaAs/AlGaAs heterostructures; ZnO/MgZnO heterostructures; AlAs quantum wells \cite{chitra2005zero,knighton2018evidence,falson2022competing,hossain2022anisotropic,madathil2023moving}, where the long-range disorder potential due to background impurities is present, and the corresponding critical density is appreciably lower.  As inferred from both the level and character of the disorder potential, Si MOSFETs and unprecedentedly high-mobility heterostructures, including SiGe/Si/SiGe heterostructures, belong to different classes of electron systems.  Importantly, the obtained contradictory results indicate that the generality of the effect observed in Si MOSFETs \cite{brussarski2018transport} is uncertain and highlight a principal problem whether one can detect a conclusive transport signature of a Wigner solid in a class of the purest accessible semiconductor heterostructures.  The problem has been solved in this paper.  Notably, its solution has required an elaboration of a complicated design of triple-gate samples.

Here, we report the observation of two-threshold $V$-$I$ characteristics accompanied by a peak of broadband current noise with frequency dependence $1/f^\alpha$ with $\alpha\leq1$ between the two threshold voltages in the insulating state at low densities ($r_{\rm s}>20$) in the 2D electron system in ultra-high mobility SiGe/Si/SiGe heterostructures.  In contrast to the standard set-up to set a current through the sample and measure the voltage, we applied a voltage and measured the current, which facilitated measurements and interpretation of the data.  The observed $V$-$I$ characteristics are very similar to those known for the collective depinning of the vortex lattice in type-II superconductors (see, \textit{e.g}., Refs.~\cite{yeh1991flux,blatter1994vortices,bullard2008vortex}), with $I$ and $V$ axes interchanged.  The results can be described by a phenomenological theory of the collective depinning of elastic structures, which naturally generates a peak of broadband current noise between the dynamic and static thresholds, the solid sliding as a whole over a pinning barrier above the static threshold.  Our results give evidence for the formation of a quantum electron solid in this electron system and show the generality of the effect for different classes of electron systems.

Data were obtained on ultra-high mobility SiGe/Si/SiGe quantum wells similar to those described in Refs.~\cite{melnikov2015ultra,dolgopolov2021valley}.  The low-temperature electron mobility in these samples reaches $\approx200$~m$^2$/Vs.  The $\approx15$~nm wide silicon (001) quantum well is sandwiched between Si$_{0.8}$Ge$_{0.2}$ potential barriers (Fig.~\ref{fig1}).  Contacts to the 2D layer consisted of $\approx 300$~nm Au$_{0.99}$Sb$_{0.01}$ alloy deposited in a thermal evaporator and then annealed.  The samples were patterned in Hall-bar shapes with the distance between the potential probes of 100~$\mu$m and width of 50~$\mu$m using photo-lithography.  An $\approx200$-nm-thick SiO layer was deposited on the surface of the wafer in a thermal evaporator, and an $\approx60$-nm-thick NiCr/Al gate was deposited on top of SiO.  After that, the contact gate was fabricated, for which the structure was covered by an $\approx150$-nm-thick SiO layer, and an $\approx40$-nm-thick aluminum gate was deposited on top of SiO.  The contact gate allowed maintaining high electron density $\approx2\times10^{11}$~cm$^{-2}$ near the contacts regardless of its value in the main part of the sample.  No additional doping was used, and the electron density was controlled by applying a positive dc voltage to the gate relative to the contacts.  The shunting channel between the contacts in the ungated area outside the Hall bar, which can reveal itself at the lowest electron densities in the insulating regime, was depleted using an additional Al gate that was fabricated simultaneously with the contact gate.  Measurements were carried out in an Oxford TLM-400 dilution refrigerator.  In the main part of the experiments, the voltage was applied between the source and the nearest potential probe over a distance of 25~$\mu$m.  The current and noise were measured by a current-voltage converter connected to a lock-in and a digital voltmeter.  The voltage-current curves were a little asymmetric with respect to reversal of the voltage; we used the negative part plotted versus the absolute value of voltage for convenience.  The electron density was determined by Shubnikov-de~Haas oscillations in the metallic regime using a standard four-terminal lock-in technique.  To improve the quality of contacts and increase electron mobility, we used a saturating infrared illumination of the samples.  The contact resistances were below 10~kOhm.  Experiments were performed on three samples, and the obtained results were similar.

\begin{figure}
\scalebox{.73}{\includegraphics[width=\columnwidth]{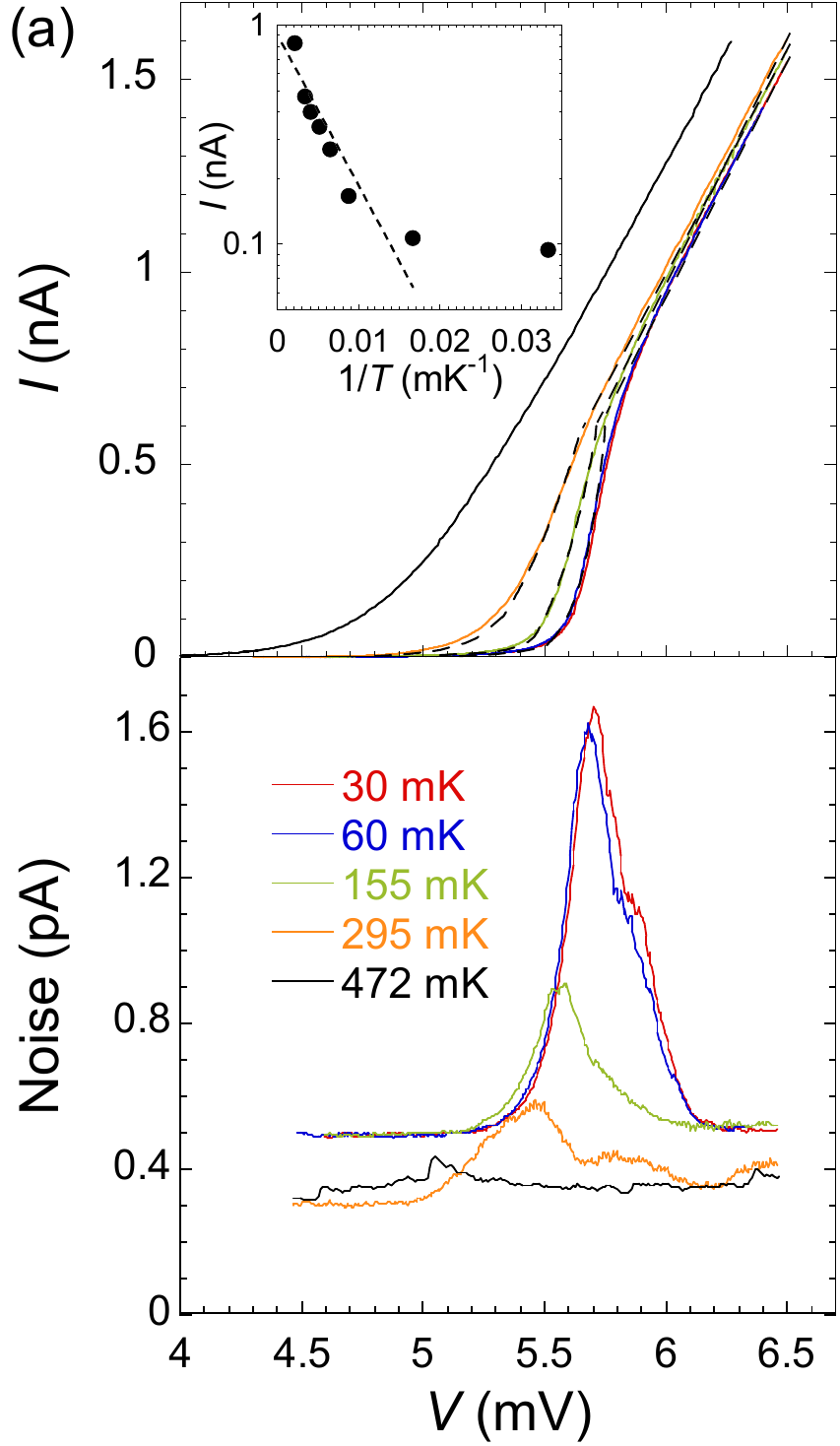}}
\hspace*{2.5mm}\scalebox{.73}{\includegraphics[width=\columnwidth]{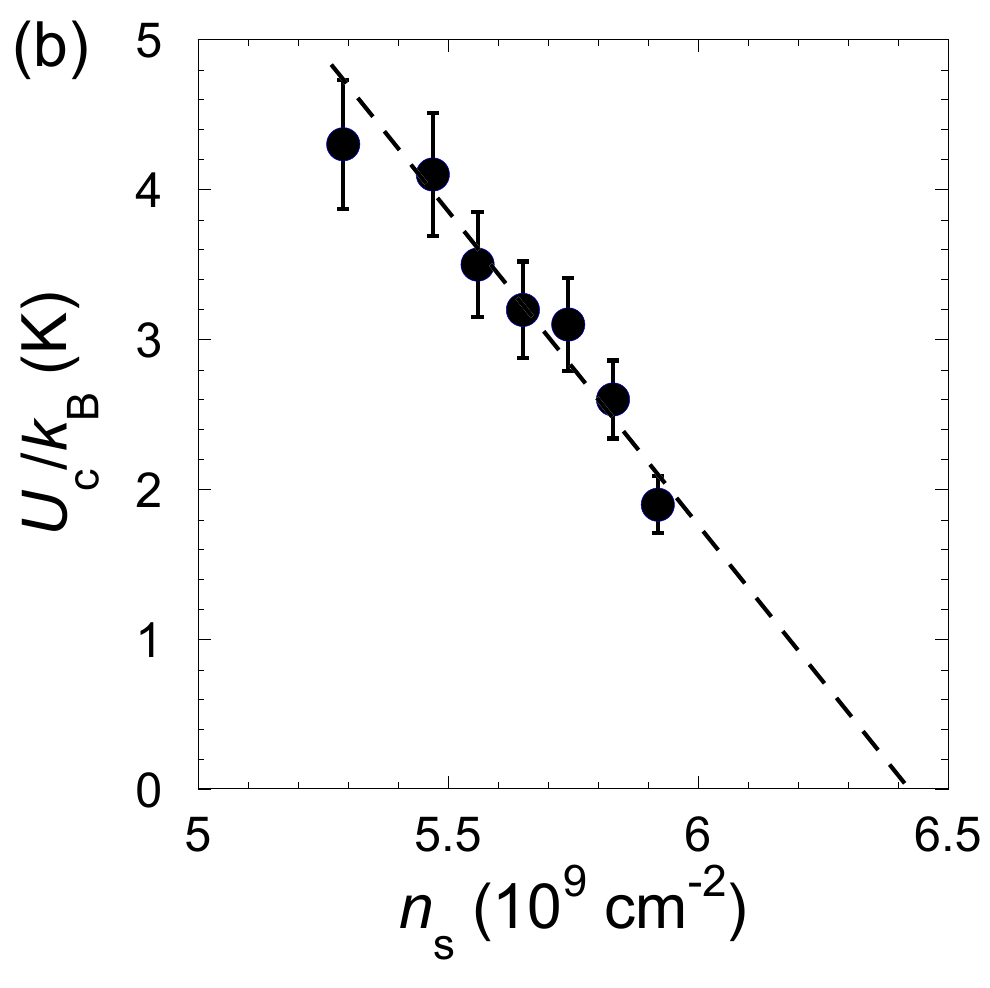}}
\caption{(a) Current (upper panel) and the broadband noise (lower panel) as a function of voltage at $n_{\rm s}=5.83\times10^9$~cm$^{-2}$ at different temperatures in sample 2.  Curves in the upper panel color-correspond to the same temperatures as in the lower panel.  The overall noise is measured over the same frequency range as that in Fig.~\ref{fig4}.  The dashed lines are fits to the data using Eqs.~(\ref{1}), (\ref{2}).  The inset shows an Arrhenius plot of $I(T)$ at $n_{\rm s}=5.83\times10^9$~cm$^{-2}$ and $V=5.6$~mV.  The dashed line is a linear fit excluding the data point at 30 mK.  (b) Activation energy $U_{\rm c}$ as a function of the electron density in sample 1.  The dashed line is a linear fit.}
\label{fig3}
\end{figure}

\begin{figure}
\scalebox{.73}{\includegraphics[width=\columnwidth]{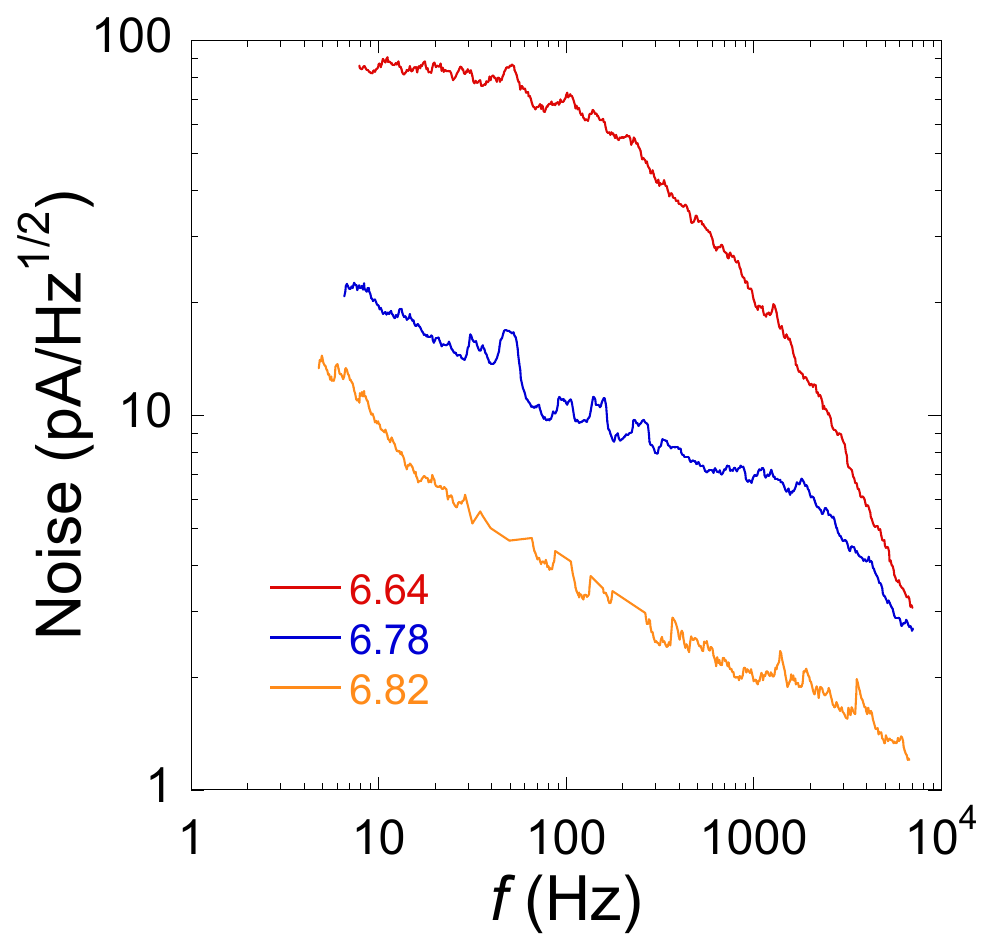}}
\caption{The frequency dependence of noise at the maximum at three electron densities indicated in units of $10^9$~cm$^{-2}$ and at $T\approx30$~mK in sample 3.}
\label{fig4}
\end{figure}

In Fig.~\ref{fig2}~(a), we show typical voltage-current characteristics (upper panel) and generated noise (lower panel) at $T\approx30$~mK at different electron densities in the insulating regime below, but not too close to $n_{\rm c}$, where $n_{\rm c}\approx8.8\times10^9$~cm$^{-2}$ is the critical density for the metal-insulator transition in our samples.  The interaction parameter $r_{\rm s}$ exceeds 20 at these values of $n_{\rm s}$.  With increasing applied voltage, the current stays near zero up to the first threshold voltage $V_{\rm{th1}}$, then the current sharply increases until the second threshold voltage, $V_{\rm{th2}}$, is reached (see also Fig.~\ref{fig2}~(b, c) for a better view).  Above the second threshold voltage, the slope of the $V$-$I$ curves is reduced, and the behavior becomes linear but not ohmic.  Between the two threshold voltages, a peak of broadband current noise is generated (the lower panel of Fig.~\ref{fig2}~(a)).  The fits (dashed lines) and the meaning of the dynamic ($V_{\rm d}$, obtained by the extrapolation of the linear part of the $V$-$I$ curves to zero current) and static ($V_{\rm s}=V_{\rm{th2}}$) thresholds indicated in Fig.~\ref{fig2}~(b, c) will be discussed below.

Figure~\ref{fig3}~(a) shows how the voltage-current characteristics and noise change with temperature.  With increasing temperature, the voltage-current characteristics become less steep between the two threshold voltages, and the curves shift to lower voltages until the two-threshold behavior disappears.  Correspondingly, the noise peak decreases and disappears with increasing $T$.  The inset to the upper panel of Fig.~\ref{fig3}~(a) is an Arrhenius plot of $I(T)$ showing the activation temperature dependence down to $T\approx60$~mK; the data point at the lowest $T$ deviates from the activation temperature dependence although the electron temperature reached $\approx30$~mK as inferred from the analysis of the Shubnikov-de~Haas oscillations in the metallic regime.  We attribute this deviation to the effect of residual sample inhomogeneities, although the explanation in terms of possible overheating effects cannot be excluded in this experiment.

In Fig.~\ref{fig4}, we show noise spectra at the maximum of noise at three electron densities and $T\approx30$~mK.  At high frequencies, the dependence of the noise on frequency becomes stronger with decreasing electron density and is close to the $1/f^\alpha$ dependence with $\alpha\approx1$ at the two lowest densities.  With decreasing frequency, the dependence at the two lowest densities changes to $1/f^\alpha$ with $\alpha\approx0.2$ which is similar to the dependence at the highest electron density.  The change of $\alpha$ occurs at a frequency that decreases with decreasing electron density.

The two-threshold $V$-$I$ characteristics reported in this paper are strikingly similar to the two-threshold $I$-$V$ characteristics known for the collective depinning of the vortex lattice in type-II superconductors, with voltage and current axes interchanged.  A phenomenological theory of the collective depinning of elastic structures was adapted for an electron solid in Ref.~\cite{brussarski2018transport}, and we briefly review it here.  With increasing voltage, the depinning of the electron solid is indicated by the appearance of a non-zero current.  Between the dynamic ($V_{\rm d}$) and static ($V_{\rm s}$) thresholds, the collective pinning of the electron solid occurs and the transport is thermally activated:
\begin{equation}
I=\sigma_0\,(V-V_{\rm d})\,\exp\left[-\frac{U_{\rm c}(1-V/V_{\rm s})}{k_{\rm B}T}\right],\label{1}
\end{equation}
where $U_{\rm c}$ is the maximal activation energy of the pinning centers, and $\sigma_0$ is a coefficient.  At voltages exceeding $V_{\rm s}=V_{\rm{th2}}$, the solid slides with friction, as determined by the balance of the electric, pinning, and friction forces, which yields the following equation
\begin{equation}
I=\sigma_0\,(V-V_{\rm d}),\label{2}
\end{equation}
where $V_{\rm d}$ corresponds to the pinning force.  Fits using these equations are shown by the dashed lines in Figs.~\ref{fig2}~(b, c) and \ref{fig3}~(a).  These describe well the experimental two-threshold $V$-$I$ characteristics.  Fitting the experimental data by Eqs.~(\ref{1}), (\ref{2}) at temperatures above $T\approx60$~mK, at which the activation temperature dependence of the current holds, allows one to extract the activation energy $U_{\rm c}$ plotted in Fig.~\ref{fig3}~(b) as a function of the electron density.  The decrease of the activation energy $U_{\rm c}$ with electron density can be described by an approximately linear dependence that tends to zero at $n_{\rm s}\approx6.4\times10^9$~cm$^{-2}$. This value is significantly lower than the critical density $n_{\rm c}$ at which the activation energy of electron-hole pairs, determined by measurements of the transport in the linear response regime, vanishes.  The coefficient $\sigma_0$, corresponding to the slope of the linear part of the $V$-$I$ curves, is approximately constant ($\sigma_0\approx1.5\times10^{-6}$~Ohm$^{-1}$) in our samples.

The observed noise peak is a natural consequence of the collective depinning of elastic structures.  Between the dynamic and static thresholds, in the regime of the collective pinning, the solid locally deforms when the depinning occurs at a particular pinning center.  Subsequently, this deformation repeats at other centers.  This generates a strong noise, as seen in Figs.~\ref{fig2}~(a) and \ref{fig3}~(a).  In contrast, above the static threshold, the solid slides as a whole over a pinning barrier, and the noise is suppressed.  As the temperature increases in the regime of the collective pinning, one approaches the regime of over-barrier sliding in which the noise is suppressed, \textit{i.e.}, the noise decreases with temperature, as seen in Fig.~\ref{fig3}~(a).

Two-threshold voltage-current characteristics very similar to those observed in our experiments were also obtained numerically in a classical model of Refs.~\cite{reichhardt2023nonlinear,reichhardt2023noise} where the driven dynamics of Wigner crystals interacting with random disorder was considered.  Between the dynamic and static thresholds, in the fluctuating regime, the noise power reaches a peak value, and the noise has a $1/f^{2\alpha}$ character.  Particularly, the frequency dependence of noise with $2\alpha=2$ at high frequencies changes to a weaker dependence with $2\alpha=0.7$ at low frequencies, which is in reasonable agreement with the experimental results.  Although the classical model \cite{reichhardt2023nonlinear,reichhardt2023noise} is used, it captures the features observed in the experiment, which adds confidence in our conclusions.  Note that somewhat smaller value of $r_{\rm s}$ for the formation of the Wigner solid compared to the predicted one can be due to the presence of a residual disorder that leads to an increase of the crystallization electron density (see, \textit{e.g.}, Ref.~\cite{chui1995impurity}).

In summary, we have observed two-threshold voltage-current characteristics accompanied by a peak of broadband current noise between the two threshold voltages in the 2D electron system in ultra-high mobility SiGe/Si/SiGe heterostructures.  The observed results can be described by a phenomenological theory of the collective depinning of elastic structures, which naturally generates a peak of a broadband current noise between the dynamic and static thresholds and changes to sliding of the solid over a pinning barrier above the static threshold.  Our results give evidence for the formation of a quantum electron solid in this electron system and show the generality of the effect for different classes of electron systems.

We are indebted to Don Heiman and Nathan Israeloff for useful discussions and critical reading of our manuscript.  The ISSP group was supported by the RF State Task.  The NTU group acknowledges support by the Ministry of Science and Technology, Taiwan (Project No.\ 112-2218-E-002-024-MBK).  S.V.K. was supported by NSF Grant No.\ 1904024.

%\bibliography{references}
%\end{document}

\end{document}